\documentclass{aa}

\usepackage{times}
\usepackage{astro}
\usepackage{aabib}

\begin{document}

\thesaurus{11(11.04.1;11.05.2;11.19.3;13.09.1)}

\title{Discovery of distant high luminosity infrared galaxies}

\author{Paul P.\ van der Werf\inst{1}
 \and D.L.\ Clements\inst{2,3}
 \and P.A.\ Shaver\inst{2}
 \and M.R.S.\ Hawkins\inst{4}}

\offprints{Paul van der Werf}
\mail{pvdwerf@strw.leidenuniv.nl}

\institute{$^1$ Leiden Observatory, P.O.\ Box 9513, NL - 2300~RA Leiden, The
Netherlands\\
$^2$ European Southern Observatory, Karl-Scharzschild-Str.~2, D-85748 
Garching bei M\"unchen, Germany\\
$^3$ Institut d'Astrophysique Spatiale, Universit\'e Paris XI,
Batiment 121, F-91405 Orsay Cedex, France\\
$^4$ Royal Observatory, Blackford Hill, Edinburgh EH9 3HJ, Scotland
}

\date{Received ... / Accepted ...}

\maketitle 

\begin{abstract}
We have developed a method for selecting the most luminous
galaxies detected by IRAS based on their extreme values of $R$, the
ratio of $60\mum$ and $B$-band luminosity.  These objects have optical
counterparts that are close to or below the limits of Schmidt surveys.
We have tested our method on a $1079\,{\rm deg}^2$ region of sky,
where we have selected a sample of IRAS sources with $60\mum$ flux
densities greater than $0.2\un{Jy}$, corresponding to a redshift limit
$z\sim1$ for objects with far-IR luminosities of $10^{13}\Lsun$. 
Optical identifications for these
were obtained from the UK Schmidt Telescope plates, using the likelihood ratio
method. Optical spectroscopy has been carried out to reliably identify
and measure the redshifts of six objects with very faint optical
counterparts, which are the only objects with $R>100$ in the sample. 
One object is a hyperluminous infrared galaxy (HyLIG) at $z=0.834$.
Of the remaining, fainter objects, five are ultraluminous
infrared galaxies (ULIGs) with a mean redshift of 0.45, higher than
the highest known redshift of any non-hyperluminous ULIG prior to this study. 
High excitation lines reveal the presence of an active nucleus
in the HyLIG, just as in the other known infrared-selected
HyLIGs. In contrast, no high excitation lines are found in the
non-hyperluminous 
ULIGs. We discuss the implications of our results for the number
density of
HyLIGs at $z<1$ and for the evolution of the infrared galaxy
population out to this redshift, and show that substantial
evolution is indicated. Our selection method is robust
against the presence of gravitational lensing if the optical and
infrared magnification factors are similar, and we suggest a way of
using it to select candidate gravitationally lensed infrared galaxies.


\end{abstract}

\section{Introduction}

One of the most fundamental results of the IRAS survey was the
discovery of a class of galaxies that emit the bulk of their energy at
far-infrared wavelengths and that have $8-1000\mum$ luminosities
$\qu{L}{IR}>10^{12}\Lsun$ (for $H_0=75\kms\pun{Mpc}{-1}$ and
$q_0=0.1$, as assumed throughout this paper).  As shown by Soifer et
al.\ (1986, 1987), these {\em ultraluminous infrared galaxies\/}
(ULIGs) are a very significant population in the local universe,
dominating the high luminosity end of the local luminosity function,
and outnumbering local optically selected quasars by a factor of at
least 2 (see \citebare{SandersMirabel96} and references therein). More
recently, a number of objects with $\qu{L}{IR}>10^{13}\Lsun$ have been
identified, for which the term {\em hyperluminous infrared galaxies\/}
(HyLIGs) has been proposed.  The first object of this type to be
discovered in an infrared-selected sample was
\object{$\sou{IRAS}{F10214{+}4724}$} at $z=2.28$
\cite{RowanRobinsonetal91a}.  Although this object is now known to be
gravitationally lensed (\citebare{BroadhurstLehar95};
\citebare{Closeetal95}; \citebare{Serjeantetal95};
\citebare{Eisenhardtetal96}), it is a HyLIG even after the
gravitational magnification has been accounted for
\cite{Downesetal95}. Furthermore, of the now more than 30 known objects in
this class, most do not show any evidence of gravitational lensing.
\nocite{Soiferetal86} 
\nocite{Soiferetal87}

The nature of the energy source powering the high
bolometric luminosities of ULIGs and HyLIGs is a subject of intense
debate (see e.g., \citebare{RowanRobinson96} for a recent review). 
Since ULIGs and HyLIGs have luminosities in the range of the most 
powerful active galactic nuclei (AGNs), it has 
been proposed that they are powered by dust-embedded AGNs. 
Alternatively, the far-infrared (FIR) luminosity may
be provided by a burst of intense star formation, with implied star formation
rates $\Mstardot\approx10^2-10^3\Msun\pun{yr}{-1}$, or even
higher for HyLIGs.
At the high end of this interval,
a starburst lasting a typical $10^8\un{yrs}$ would produce
a stellar mass of $\sim10^{11}\Msun$, comparable to the luminous mass of a
typical galaxy. Therefore, if powered only by star formation, the 
HyLIGs form stars at the rate expected for high redshift galaxies in their
formation process, undergoing an initial starburst that builds up the
bulk of their stellar population. 
AGNs are known to be present in most of the known HyLIGs, but
since the fraction of the FIR luminosity that they provide is unknown,
the presence of intense star formation is not ruled out. Indeed, several lines
of evidence indicate the presence of vigorous star formation in 
\object{$\sou{IRAS}{F10214+4724}$} 
\cite{RowanRobinsonetal93,Krokeretal96,GreenRowanRobinson96}, in addition to 
an embedded AGN \cite{Elstonetal94,Soiferetal95,Goodrichetal96}. Furthermore,
since most of the known HyLIGs have been found in
surveys of AGNs, the
ubiquity of AGNs in known HyLIGs may be entirely a selection effect. Thus
an unbiased survey for HyLIGs based on a sample of faint
far-IR sources would be valuable for a reliable
assessment of the nature of these objects. In addition, since HyLIGs can be
observed to very significant redshifts, a determination of their number 
densities will strongly constrain the evolution of IRAS galaxies at these 
redshifts, as far as the most luminous part of the luminosity function is
concerned. 

In this paper we present a survey for the most luminous galaxies 
detected by IRAS in a $1079\,{\rm deg}^2$ area of sky. 
We discuss the sample selection process in 
\secref{sec.sel} and our observation and reduction procedures in 
\secref{sec.obs}. The results are presented and discussed in 
\twosecsref{sec.results}{sec.discussion}, 
while our conclusions are summarized in 
\secref{sec.conclusions}.

\section{Sample selection}\label{sec.sel}

We based our survey on a sample of sources selected from the IRAS
Faint Source Catalog (FSC; \citebare{Moshiretal92}). Since our project
involves the optical identification of faint sources, it is important
to avoid spurious FSC sources (resulting from e.g., small-scale
structure in $60\mum$ cirrus). We therefore selected a survey area
where diffuse $60\mum$ emission is faint (cf., the IRAS $60\mum$ maps
presented by \citetext{RowanRobinsonetal91b}).  The area selected
consists of the 4~hour\break 
R.A.~(B1950.0) interval between $21^{\rm h}$
and $1^{\rm h}$, at Dec.~(B1950.0) less than $-30\deg$ and Galactic
latitude less than $-40\deg$. We also used a $60\mum$ flux cutoff of
$0.2\un{Jy}$, since below this value the FSC becomes rapidly less
complete.  The region selected contains 3057 $60\mum$ FSC sources
brighter than $0.2\un{Jy}$ over an area of $1079\,{\rm deg}^2$.  Stars
and nearby galaxies were rejected by excluding all sources detected at
$12\mum$, leaving 2719 objects in the sample.  As a further safeguard
against spurious sources, we used the FSC flux quality indicators and
cirrus flag, to retain in the sample only those sources with a
high-quality detection at $60\mum$ and no confusion by cirrus.  Since
the spectral energy distribution (SED) of ULIGs peaks in the
rest-frame $60\mum$ region, ULIGs at $z>0.3$ will have flux densities
rising monotonically with wavelength in the IRAS bands. Therefore the
resulting sample was further reduced by retaining only sources
detected at both 60 and $100\mum$. However, following
\citetext{Clementsetal96}, sources with $S_{100}/S_{60}>5$ (where
$S_{100}$ and $S_{60}$ are the 100 and $60\mum$ flux densities as
given in the FSC) were excluded, since such cold sources most likely
arise from small-scale structure in Galactic cirrus.  Finally, we
rejected sources with associations in other catalogs as indicated in
the FSC, thus limiting our sample to 313 objects. As shown in
\secref{sec.discussion}, these strict selection criteria make our
survey a sparse (approximately 1 in 8) but unbiased survey for
infrared galaxies with $S_{60}\ge0.2\un{Jy}$ over the
$1079\pun{deg}{2}$ survey area.

In order to select from our sample the most distant objects, we define 
the FIR loudness $R$ by
\begin{equation}\label{eq.R}
R\equiv L_{60}/L_B=S_{60}\,10^{0.4(B-14.45)},
\end{equation}
(see \citebare{Clementsetal96}), where $L_{60}$ is the $60\mum$ luminosity, 
$L_B$ the luminosity in the $B$ band, $B$ the $B$-band magnitude, and $S_{60}$ 
the $60\mum$ flux density in Jy.
The bivariate $B$-$60\mum$ luminosity function has been derived by
\citetext{Saundersetal90}, who show that $R$ increases monotonically
with $L_{60}$. This dependence accounts for the fact that
the high luminosity cutoff of the luminosity function is much sharper
in the optical regime than in the infrared. Therefore, $S_{60}$ can
be combined with the apparent $B$ band magnitude to calculate $R$ and hence 
obtain a crude
estimate of the far-IR luminosity and distance of the object. An approximate
$B$ magnitude of the most luminous sources in our sample
can be estimated as follows. For the cosmological parameters adopted
here, a ULIG with
$\qu{L}{IR}=10^{12}\Lsun$ will have $S_{60}=0.2\un{Jy}$ if it is at
$z\approx0.35$. Using the bivariate luminosity function of
\citetext{Saundersetal90}, 95\% of these will have $R>10$, and they will have 
a mean absolute $B$
magnitude of $\mgd -20.0$, or an apparent magnitude $B\approx\mgd 21.0$ at
$z\approx0.35$. At these magnitudes, sources can be identified on
optical Schmidt survey plates. Furthermore, since the IRAS
error ellipse for our sample sources is typically $10''\times30''$,
and the extragalactic source density at $B<21\mg$ is about 1000
per square degree, there is only about a 2\% probability of chance
superpositions at these magnitudes. Therefore, ULIGs in the FSC can
be identified in optical Schmidt surveys and selected based on their
$L_{60}$/$L_B$ ratio. This method has been succesfully used by 
\citetext{Clementsetal96}, to find 91 ULIGs with a median redshift
between 0.2 and 0.3, and a maximum redshift of 0.43, by selecting only 
$R>10$ objects from FSC 
sources identified on Schmidt plates. This reasoning suggests that HyLIGs 
could be found in the same way, but selecting only sources with
$R>100$ (e.g., \object{$\sou{IRAS}{F10214{+}4724}$} has $R=350$).
However, a HyLIG with $\qu{L}{IR}=10^{13}\Lsun$ will have
$S_{60}=0.2\un{Jy}$ at $z\approx1$, and a most likely
$B\approx\mgd 23.7$. Such sources are below the plate limit of common
Schmidt surveys, while with deeper imaging the density of faint sources becomes
so high, that reliable identification is no longer possible without additional
information. In order to circumvent these problems we have first
obtained a sample of {\em candidate\/}
distant HyLIGs by selecting
those sources for which no reliable identification can be found on
optical Schmidt plates, or which have very faint optical counterparts. Optical
follow-up (see \secref{sec.obs}) was then used to obtain the correct 
identifications and redshifts for the candidates.
We note that the 
existence of faint but reliable FSC sources without optical counterparts above
the typically $B_J=22\mg$ limit of Schmidt surveys has been noted by 
several groups 
(\citebare{Wolstencroftetal86}; \citebare{RowanRobinson91}; 
\citebare{Sutherlandetal91}; \citebare{Clementsetal96}; 
\citebare{Oliveretal96a}). Our programme 
is the first published project to systematically identify these
optically faint and potentially very distant sources.

We carried out an identification programme for our FSC sources on the
U.K.\ Schmidt Telescope southern sky survey plates, digitized
using the COSMOS plate scanning machine \cite{Yentisetal92}. The
COSMOS catalog provides $B_J$ magnitudes to a completeness limit of
$B_J\approx 22\mg$, positions, major and minor axis lengths, position
angles, and for objects with
$B_J<21\mg$ a classification as star or galaxy, 
based on an algorithm described by 
\citetext{HeydonDumbletonetal89}. The identifications were performed
using the likelihood ratio method (see \citetext{SutherlandSaunders92}
for a detailed discussion). Briefly, the method assigns to every
optical source a likelihood ratio
\begin{equation}
L={e^{-r^2/2}\over2\pi\sigma_1\sigma_2 N({<}B_J),}\label{eq.L}
\end{equation}
where $r$ is the distance between the FSC and optical positions in a
coordinate system where the IRAS error ellipse is a circle of unity
radius,
\begin{equation}
r=\sqrt{\left(d_1\over\sigma_1\right)^2+\left(d_2\over\sigma_2\right)^2}.
\end{equation}
In these expressions, $\sigma_1$ and $\sigma_2$ are the major and
minor axes of the FSC error ellipse, $d_1$ and $d_2$ are the position
differences of the optical source with respect to the FSC source
projected on these axes, and $N({<}B_J)$ is the density of objects
brighter than the candidate object. 

In calculating $L$, it is important to take into account possible
errors in the star/galaxy classification performed by\break
COSMOS\null. 
We noted that a number of FSC sources
had counterparts with a very high value of $L$, which were however
classified by COSMOS as stellar, but with axial ratios significantly
exceeding unity. During our observing programme described in
\secref{sec.obs} we obtained $B$-band images of 19 of these, covering
a range of magnitudes and axial ratios. 
In these 19 fields, we found that all objects
classified as stellar by COSMOS, were in fact galaxies if they had
$B_J<18\mg$ and axial ratio exceeding 1.27. Some objects with lower
axial ratios were also misclassified as stellar.
We therefore reclassified all objects with $B_J<18$ and axial ratio
greater than 1.27 as galaxies.  
In calculating $L$ for objects classified as galaxies, 
we computed $N({<}B_J)$ taking into
account only objects having the same classification. For objects
classified as
stellar or having $B_J>\mgd 20.5$ (making them too faint for useful 
classification), the calculation of $N({<}B_J)$ took into account all
sources, regardless of classification. This method allows for the
possibility of misclassification, while still somewhat favouring
objects classified as galaxies.

In \eqref{eq.L} the simplifying assumption is made that the
position errors in the FSC are Gaussian. This assumption is
approximately correct for small $r$, but the FSC position error
distribution has wings which are stronger than Gaussian ones
\cite{SutherlandSaunders92,Clementsetal96,Bertinetal97}. In order to
take these wings into account, all optical sources classified as galaxies were
assigned $L=5$ if they were within $1'$ from the FSC position and had
$B_J<19\mg$. 

The identification process consisted of calculating $L$ as described
above for every
optical object within $2'$ of the FSC position. 
Following \citetext{Clementsetal96}, we consider an optical
identification reliable for $L\ge5$. Of our sample of 313 FSC sources,
302 had identifications with $L\ge5$. Of the remaining 11, 5 were found
to have $B_J<\mgd19.5$ objects within $\mind 1.25$ of the FSC positions,
which present plausible identifications given the non-Gaussian wings
of the FSC position error distribution. The remaining 6 had no
plausible optical counterpart with $B_J<\mgd20.5$. The best ``identifications''
for these FSC sources had $L<2$. 
These 6 sources thus form our sample of
candidate HyLIGs. We note that \object{$\sou{IRAS}{F10214{+}4724}$}, 
which has $B=\mgd 22.5$ and is located outside the IRAS FSC error
ellipse, would also have been selected by this method,
if it was located within our survey area.

\section{Observations and reduction}\label{sec.obs}

In order to obtain identifications and redshifts of our FSC\break 
sources we
used the ESO Faint Object Spectrograph and Camera (EFOSC;
\citebare{Buzzonietal84}) on
the $3.6\un{m}$ telescope of the European Southern Observatory at La
Silla, Chile, on the nights of September 6--8, 1994. Conditions were
clear but not photometric and the seeing (measured in Gunn $i$-band)
was typically $\secd 1.5$. The detector was
a $512^2$ anti-reflection coated, thinned, back-illuminated Tektronix
CCD with a pixel size of $\secd 0.61$ in imaging mode. 
Our observing strategy was as follows.
First, a short ($1-2\un{min}$) exposure of the field was taken in
imaging mode using a
$B$-band filter (and for some objects also in $R$ and/or Gunn $i$),
allowing the detection of objects several magnitudes below the COSMOS
plate limit. Subsequently, long-slit spectra were taken of potential
counterpart galaxies, in order of decreasing
likelihood ratio
$L$, and including galaxies above and below the COSMOS plate limit,
until an emission line galaxy was found.
The spectra were taken with integration times of 10 to $30\un{min}$, 
using the B300 and R300 grisms with a $\secd 1.5$ slit, providing a spectral
resolution $\lambda/\Delta\lambda\approx300$ from 3640 to $6860\un{\AA}$
(B300) or from 5970 to $9770\un{\AA}$ (R300). Wavelength calibration was
derived from exposures of a HeAr lamp. Photometric and spectroscopic
calibration was achieved by observations of the spectrophotometric
standard L$870{-}2$. Flatfields were obtained from lamp exposures.
Data reduction was performed using the standard long-slit reduction
procedures as implemented in the IRAF package.

\begin{table*}
\caption[]{Parameters of distant FSC sample sources\label{tab.results}}
\begin{flushleft}
\begin{tabular}{ccccccccr}
\noalign{\smallskip}
\hline
\noalign{\smallskip}
name & R.A. & Dec. & $z$ & $S_{60}$ & $S_{100}$ & $B_J$ & $\qu{L}{FIR}$ & $R$\\
           & \multicolumn{2}{c}{(B1950.0)} & & [Jy] & [Jy] & & [$\Lsun$] & \\
\noalign{\smallskip}
\hline
\noalign{\smallskip}
\object{$\sou{IRAS}{F00320{-}3307}$} & $\hmsd 00h32m00.99s$ 
& $\dmsd -33d07m44.6s$ & 0.439 & 0.43 & 0.87 & $\mgd 21.10$ 
& $4.9\times10^{12}$ & 200 \\
\object{$\sou{IRAS}{F00417{-}3358}$} & $\hmsd 00h41m40.41s$ 
& $\dmsd -33d58m03.0s$ & 0.461 & 0.24 & 0.59 & $\mgd 21.97$ 
& $3.3\times10^{12}$ & 240 \\
\object{$\sou{IRAS}{F21065{-}3451}$} & $\hmsd 21h06m33.93s$ 
& $\dmsd -34d51m34.0s$ & 0.329 & 0.36 & 1.55 & $\mgd 21.54$ 
& $3.1\times10^{12}$ & 250 \\
\object{$\sou{IRAS}{F21243{-}4501}$} & $\hmsd 21h24m26.05s$ 
& $\dmsd -45d01m54.2s$ & 0.834 & 0.30 & 0.90 & $\mgd 23.7\hphantom{0}$ 
& $1.9\times10^{13}$ & 1500 \\
\object{$\sou{IRAS}{F22148{-}4013}$} & $\hmsd 22h14m52.07s$ 
& $\dmsd -40d12m58.4s$ 
& 0.529 & 0.33 & 0.83 & $\mgd 21.6\hphantom{0}$ & $4.4\times10^{12}$ & 240 \\
or:            & $\hmsd 22h14m51.96s$ & $\dmsd -40d13m02.7s$ & 0.380 &
     &      & $\mgd 21.5\hphantom{0}$ & $2.0\times10^{12}$ & 220 \\
\object{$\sou{IRAS}{F23555{-}3436}$} & $\hmsd 23h55m32.14s$ 
& $\dmsd -34d36m29.3s$ & 0.490 & 0.31 & 0.71 & $\mgd 20.98$ 
& $4.8\times10^{12}$ & 130 \\
\noalign{\smallskip}
\hline
\noalign{\smallskip}
\end{tabular}
\end{flushleft}
\end{table*}

\section{Results}\label{sec.results}

Our results are summarized in \tabref{tab.results}. Positions refer to
the positions provided by the COSMOS catalog, or for sources below the
COSMOS plate limit to detections on our EFOSC images, adopting the astrometry
of the COSMOS plates. Position errors are about $1''$
r.m.s. Magnitudes $B_J$ are values provided by COSMOS, except for
\object{$\sou{IRAS}{F21243{-}4501}$}, 
where magnitudes are derived from our EFOSC
imaging, calibrated using the COSMOS $B_J$
magnitudes of other sources in the field. Since this procedure ignores the 
differences between $B_J$ and the Bessel $B$ filter used in EFOSC, 
$B_J$ for this
object is only approximate. Far-IR
luminosities $\qu{L}{FIR}$ have been calculated from
$\qu{L}{FIR}=4\pi\qu{D}{L}^2\qu{F}{FIR}$ where $\qu{D}{L}$ is the
luminosity distance and 
\begin{equation}\label{eq.FFIR}
{\qu{F}{FIR}\over\esc}=
1.8\times10^{-11}\left(2.58{S_{60}\over\un{Jy}}+{S_{100}\over\un{Jy}}\right)
\end{equation}
(see \citebare{SandersMirabel96} and references therein).
This procedure somewhat underestimes the total far-IR luminosity because
it does not include a $K$-correction. An accurate\break 
$K$-correction
is not possible because of the lack of knowledge of the SED of the
sources. However, under the assumption that the SED is similar to that
of the prototypical ULIG \object{$\sou{Arp}{220}$}, we find that the
underestimate introduced by \eqref{eq.FFIR} could be up to 50\% for
the most distant objects.

Notes on individual sample sources:
\begin{itemize}
\item \object{$\sou{IRAS}{F00320{-}3307}$}: 
while classified as only one galaxy by
COSMOS, this system consists of 2 interacting galaxies at $z=0.439$. 
The compound
spectrum shows strong [$\ion{O}{ii}$] $3727\un{\AA}$, in addition to
[$\ion{Ne}{iii}$] $3869\un{\AA}$, [$\ion{O}{iii}$] $5007\un{\AA}$ and
$\Ha$.
\item \object{$\sou{IRAS}{F00417{-}3358}$}: the object with the highest
likelihood ratio in this field, and therefore the a priori
most likely counterpart, was found to be a luminous object showing
[$\ion{O}{ii}$], [$\ion{Ne}{iii}$], Ca H and K absorption, and a
$4000\un{\AA}$ break at $z=0.461$.
\item \object{$\sou{IRAS}{F21065{-}3451}$}: the second most likely counterpart
as indicated by our identification process, barely resolved at $z=0.329$, and
showing strong [$\ion{O}{ii}$], in addition to $\rec{H}{b}$ and
[$\ion{O}{iii}$].
\item \object{$\sou{IRAS}{F21243{-}4501}$}: none of the possible COSMOS
identifications showed emission lines, but a fainter object close to
the IRAS error ellipse was found to have [$\ion{Ne}{v}$] $3426\un{\AA}$
and [$\ion{O}{ii}$] at $z=0.834$. A broad feature at the expected
wavelength of $\ion{Mg}{ii}$ $2798\un{\AA}$ may also be present.
\item \object{$\sou{IRAS}{F22148{-}4013}$}: spectra of two galaxies only $\secd
4.5$ apart yielded redshifts of 0.380 and 0.529, based on strong
[$\ion{O}{ii}$], $\rec{H}{b}$ and [$\ion{O}{iii}$] lines (for both
objects) and also strong $\Ha$, [$\ion{N}{ii}$] $6584\un{\AA}$ and
[$\ion{S}{ii}$] 6716 and $6731\un{\AA}$ lines (in the object at
$z=0.380$). The two objects are of closely similar $B_J$
magnitude. Either of these may be the correct identication, or they
may both contribute part of the FSC $60\mum$ flux density.  In either
case at least one of the objects is a ULIG, but none is a HyLIG\null.
\item \object{$\sou{IRAS}{F23555{-}3436}$}: the most likely identication from
the COSMOS plate is a distorted object 
outside but close to the FSC error ellipse showing
[$\ion{O}{ii}$] and [$\ion{Ne}{iii}$] at $z=0.490$.
\end{itemize}

In addition, we observed one object not in our sample
of 6 candidate distant objects, 
\object{$\sou{IRAS}{F22569{-}5523}$},
in order to check
possible misidentification, since the only likely
counterpart was classified by COSMOS
as a fairly bright star 
with low axial ratio. However, this object is in
fact a $B_J=\mgd 17.86$ galaxy with a possible tail or
extension towards the east, and showing [$\ion{Ne}{iii}$],
[$\ion{O}{ii}$], $\rec{H}{b}$ and [$\ion{O}{iii}$] emission lines at
$z=0.235$. The COSMOS position for this object is R.A.\ (1950) =  
$\hmsd 22h56m53.73s$, Dec.\ (B1950) = $\dmsd -55d23m24.6s$ and its
far-IR luminosity is $8.8\times10^{11}\Lsun$.
Accounting for the flux beyond $100\mum$, this object is
also a ULIG\null.

\section{Discussion}\label{sec.discussion}

All objects from our sample are found to have a FIR loudness
$R>100$. In contrast, the highest value of $R$ among the $L\ge5$ sources
that were removed from the sample is 76. Therefore our approach of
selecting those sources which do not have reliable counterparts above
the COSMOS plate limit, or for which the counterpart is so faint that
misidentification is no longer unlikely, proves to be very effective
in selecting sources with extreme values of $R$.  What is the nature
of these objects?  Of our sample of 6 sources, one is a HyLIG and five
are non-hyperluminous ULIGs. The five non-hyperluminous 
ULIGs are all detected on the COSMOS plates and
have $B_J<22.0$ and $R<250$.  Their {\em mean\/} redshift $z=0.45$ is
higher than the highest known redshift of any non-hyperluminous 
ULIG prior to this
study, indicating that our procedure is also a powerful method for
selecting distant ULIGs.  The HyLIG in our sample is the only object
not detected on the COSMOS plates and this object has $B=\mgd 23.7$ and
$R=1500$. This result confirms that HyLIGs can be found by selecting
objects with extreme values of $R$. The main difficulty in applying
this method is the large size of the IRAS position error ellipses,
which precludes a direct optical identification at the faint magnitude
levels expected for distant HyLIGs. However, future surveys, such as
the ongoing European Large Area Infrared Survey (ELAIS;
\citebare{Oliver96}), and surveys with SIRTF and FIRST, and with SCUBA
on the James Clerk Maxwell Telescope (JCMT) will provide substantially
better positional accuracy and not suffer from this identification
ambiguity. The method used here for selecting the most luminous and
distant objects can be adapted directly to those surveys.

The small size of our sample, which contains only one\break 
HyLIG, precludes
any detailed statistical inferences, which\break
must await more extensive
programmes using this selection and identification method, based on IRAS
data or on the surveys mentioned previously.  However, a number of
trends in our data merit further discussion.  In the first place, the
detection of
[$\ion{Ne}{v}$] emission in the only HyLIG in our sample
shows that this object contains an AGN\null.  Thus all three
IRAS-selected HyLIGs discovered so far
(\object{$\sou{IRAS}{F10214{+}4724}$},\break 
\object{$\sou{IRAS}{F15307{+}3252}$}
\cite{Cutrietal94,Hinesetal95} and \object{$\sou{IRAS}{F21243{-}4501}$} (this
work)) contain AGNs. While the\break
statistics for HyLIGs is still based on
small numbers, the result is significant, since the
[$\ion{Ne}{v}$]
line was not detected in any of the non-hyperluminous 
ULIGs in our sample, while our
spectra did cover the wavelength where this line would be expected.
Thus the HyLIGs form a remarkable contrast with the non-hyperluminous
ULIGs, where the
presence or absence of AGNs is a strongly debated issue, and direct
evidence for the presence for an AGN is very scarce.

Our procedure brings about incompleteness in our sample of $R>100$ objects 
in two ways: identification incompleteness and selection
incompleteness. The former effect arises if objects with
$R>100$ fail to be selected by our $L<5$ criterion, which occurs if a
bright galaxy lies close to the line-of-sight to a distant FSC
source, giving rise to erroneous identification with the bright galaxy. 
As noted in \secref{sec.sel}, the probability of
misidentification in this situation is only about 2\% for galaxies
with $B_J<21\mg$. Since the large majority of our $L\ge5$
identifications have counterparts significantly brighter than
$B_J=21\mg$ (for 85\% of the objects with $L\ge5$, the counterpart has
$B_J\le\mgd19.0$), the probability of chance superpositions is much
less than 2\%, and the identification incompleteness can thus be
neglected. 

However, the sample of 313 objects used for our identification programme does
suffer from selection incompleteness. Our
selection method was aimed at rejecting spurious sources; however, as
shown below, it must have
removed a significant number of real sources from the sample
as well. The relevant selection criteria are the requirement to have a
high-quality $60\mum$ detection, no cirrus confusion, and a detection
at $100\mum$. While these criteria were effective at rejecting
spurious detections, they also introduce a selection
incompleteness, and may have rejected some distant objects. 
In order to assess the magnitude of this effect, we compare our sample
to the \hbox{FSS-$z$~I} sample described by
\citetext{Oliveretal96a}. This sample has been constructed using
low-cirrus regions with good IRAS $60\mum$ coverage and is estimated
to be 99\% complete for $S_{60}\ge 0.2\un{Jy}$, which is the same flux
limit as the sample described in the present paper. It contains 1931 IRAS
FSC galaxies over an area of $839\,{\rm deg}^2$, giving a source density of 
2.30 per deg$^2$. Adopting this source density as characteristic for
the present survey shows that a
total of 2483 expected IRAS FSC galaxies over the entire
survey area should be expected, 
a plausible number given that, including spurious
sources, our initial extragalactic sample in this area contained 2719
objects (see \secref{sec.sel}).
In contrast, only 313 objects were retained in our sample of candidate objects
after the strict 
selection criteria described in \secref{sec.sel} had been applied.
However, since none of these criteria introduces a bias in
luminosity or distance, our sample is 
{\em unbiased} and our survey thus constitutes a {\em sparse\/} (approximately
1 in 8) survey of infrared galaxies with $S_{60}\ge0.2\un{Jy}$ 
over the $1079\pun{deg}{2}$ area.
Hence we can use our results to
estimate a number density for HyLIGs at $z\le1$ of approximately
$7\times10^{-3}\pun{deg}{-2}$, with considerable uncertainty due to
the small numbers involved. We note that,
adopting the local $60\mum$ luminosity function of 
\citetext{Saundersetal90}, this estimate implies significant
evolution in the infrared galaxy population to $z=1$. Only in the
unlikely case that the
HyLIG detected in our sparse survey was the only $z\le1$ HyLIG in the entire
$1079\pun{deg}{2}$ survey area, no evolution would be needed. 

We finally note that since we are using $L_{60}{/}L_B$ to select
luminous objects, our selection method is robust against the presence
of gravitational lensing, provided the corresponding magnification
factors are similar at $60\mum$ and $B$. As a result, once a redshift
and hence an infrared luminosity is available, $R$ and $\qu{L}{IR}$
may be combined to address the possibility of gravitational lensing.
We illustrate the method using the lensed HyLIG
\object{$\sou{IRAS}{F10214{+}4724}$} and the HyLIG 
\object{$\sou{IRAS}{F21243{-}4501}$},\break
identified in the present work. As noted in \secref{sec.sel},\break
\object{$\sou{IRAS}{F10214{+}4724}$} has $R=350$. Using the bivariate
$B$-$60\mum$ luminosity function of \citetext{Saundersetal90}, we find
a most likely intrinsic infrared luminosity $\qu{L}{IR}^{\rm intr}$ of
about $3\times10^{12}\Lsun$. 
The apparent luminosity following from the redshift of
2.28 on the other hand, is $\qu{L}{IR}^{\rm
app}=2\times10^{14}\Lsun$. The large
discrepancy between $\qu{L}{IR}^{\rm intr}$ and $\qu{L}{IR}^{\rm app}$
suggests gravitational amplification by a factor of about 60. 
Using the same reasoning, for
\object{$\sou{IRAS}{F21243{-}4501}$} we find 
$\qu{L}{IR}^{\rm intr}=1.6\times10^{13}\Lsun$
and $\qu{L}{IR}^{\rm app}=1.9\times10^{13}\Lsun$. 
Because of the similarity of the
two values, there is in this case no indication for gravitational
lensing. Caution is required when applying this method, since the
underlying assumption of similar magnification factors at optical and
infrared wavelengths may easily be violated, as is the case in
\object{$\sou{IRAS}{F10214{+}4724}$}, where an optical magnification
by approximately a factor of 100 is found \cite{Eisenhardtetal96},
whereas the infrared magnification is only approximately a factor of
10 (\citebare{Downesetal95}; \citebare{GreenRowanRobinson96};
\citebare{Serjeantetal98}). 
Therefore the actual
presence or absence of 
gravitational amplification must always be established by
additional observations. However
this method may be useful for selecting candidate
gravitationally lensed sources for further study.

\section{Conclusions}\label{sec.conclusions}

\begin{enumerate}
\item We have identified the most luminous infrared 
galaxies in an unbiased sample of 313
reliable extragalactic IRAS FSC sources with $S_{60}>0.2\un{Jy}$. Our
method is based on the bivariate $B$-$60\mum$ luminosity function of
infrared galaxies, which implies that the most luminous objects have
the highest values of $R$ (as defined by \eqref{eq.R}), and have
optical counterparts that are so faint that they cannot be reliably
identified (or are undetected) in typical Schmidt surveys. Using
optical spectroscopy, we have systematically identified the optical
counterparts of all of the 6 sources in our $60\mum$ sample that were
too faint in $B_J$ to be reliably identified on the UKST plates. Our
results confirm that this method selects the galaxies with the 
largest values of
$R$, so that these galaxies are indeed the 6 most luminous infrared galaxies
in our sample. Five of these are non-hyperluminous 
ULIGs with a mean redshift of 0.45,
higher than any previously known non-hyperluminous 
ULIG; the remaining source is a
HyLIG at $z=0.834$. 
\item The HyLIG in our sample (\object{$\sou{IRAS}{F21243{-}4501}$}) contains
an AGN, as shown by the presence of [$\ion{Ne}{v}$] emission. Hence
all infrared-selected HyLIGs discovered so far unambiguously show the
presence of AGNs. In contrast, none of the non-hyperluminous
ULIGs in our sample show
evidence for the presence of AGNs, and such evidence is rare among
non-hyperluminous ULIGs in general.
\item Our method is robust against the effects of gravitational
lensing if the optical and infrared magnification factors are
similar. Under this assumption this method may be useful for selecting 
candidate gravitationally lensed sources by comparing an intrinsic
luminosity (estimated from $R$) with the apparent luminosity
(calculated from $S_{60}$ and $z$). 
\item Our survey consitutes an unbiased, sparse (approximately 1 in 8)
survey of infrared galaxies with $S_{60}\ge0.2\un{Jy}$ over a
$1079\pun{deg}{2}$ area, and the results allow an estimate of the
number density of HyLIGs at
$z\le1$ of approximately\break
$7\times10^{-3}\pun{deg}{-2}$, with
considerable uncertainty due to the small numbers involved. Compared to the
local luminosity function of infrared galaxies, this estimate indicates
substantial evolution at the highest luminosities, except in the
unlikely case that 
the HyLIG found in our sparse survey is the only HyLIG at
$z\le1$ in the entire $1079\pun{deg}{2}$ survey area.
\end{enumerate}

\begin{acknowledgements}
This work was supported in part by the ``Surveys with the Infrared
Space Observatory'' network set up by the European Commission under
contract ERB FMRX-CT96-0068 of its TMR programme.
This paper is based on observations made at the European Southern 
Observatory, La Silla, Chile.
The Infra Red Astronomical Satellite (IRAS) was developed and operated
by the Netherlands Agency for Aerospace Programs (NIVR), the U.S.\
National Aeronautics and Space Administration (NASA) and the U.K.\
Science and Engineering Council (SERC).
The research of Van der Werf has been made possible by a fellowship of
the Royal Netherlands Academy of Arts and Sciences.
\end{acknowledgements}

\bibliographystyle{astrobib_aa}
\bibliography{%
strings,%
datareduction,%
farIR,%
galaxyevolution,%
HyLIGs,%
IRASF10214+4724,%
ISM,%
largescalestructure,%
optical,%
ULIGs%
}

\end{document}